# Highly accurate prediction of material optical properties based on density functional theory

Mitsutoshi Nishiwaki and Hiroyuki Fujiwara[*]

*Department of Electrical, Electronic and Computer Engineering, Gifu University, 1-1 Yanagido, Gifu 501-1193, Japan*

**Abstract**

Theoretical material investigation based on density functional theory (DFT) has been a breakthrough in the last century. Nevertheless, the optical properties calculated by DFT generally show poor agreement with experimental results particularly when the absorption-coefficient ($\alpha$) spectra in logarithmic scale are compared. In this study, we have established an alternative DFT approach (PHS method) that calculates highly accurate $\alpha$ spectra, which show remarkable agreement with experimental spectra even in logarithmic scale. In the developed method, the optical function estimated from generalized gradient approximation (GGA) using very high-density $k$ mesh is blue-shifted by incorporating the energy-scale correction by a hybrid functional and the amplitude correction by sum rule. Our simple approach enables high-precision prediction of the experimental $\alpha$ spectra of all solar-cell materials (GaAs, InP, CdTe, $CuInSe_2$ and $Cu_2ZnGeSe_4$) investigated here. The developed method is superior to conventional GGA, hybrid functional and GW methods and has clear advantages in accuracy and computational cost.

*Corresponding author, E-mail address: fujiwara@gifu-u.ac.jp

## 1. Introduction

Prediction of material optical properties based on density functional theory (DFT) has been a revolutionary technique that allows quite effective optical-material searches even without forming materials experimentally [1-4]. The DFT methods for optical-function calculation have already been established [5-9] and a vast amount of optical spectra deduced from DFT calculations have been reported [1-15]. Nevertheless, a critical view point that has been lacking in conventional DFT optical-function calculations is the justification of calculation results. In particular, calculated DFT spectra vary rather significantly with the choice of the approximation method [8-10]. Thus, the DFT calculation results need to be verified based on experimental spectra.

So far, optical-function calculations by DFT have mainly been justified by the comparison with experimental dielectric functions ($\varepsilon = \varepsilon_1 - i\varepsilon_2$) in linear scale [4-9]. In photovoltaic device simulations, however, the absorption-coefficient ($\alpha$) spectrum in logarithmic scale ($\alpha = 10^2 \sim 10^6$ cm$^{-1}$) is generally required [16,17]. When the logarithmic $\alpha$ spectra obtained from experiment and DFT calculation are compared, the agreement is generally poor and the difference between the theoretical and experimental values often reaches almost one order of magnitude [10,15]. Accordingly, there is a strong requirement for establishing a DFT calculation method that can accurately predict $\alpha$ spectra in conventional logarithmic scale.

We previously found that very-high-density $k$ mesh calculations are essential to accurately reproduce the $\alpha$ variation of various solar cell materials particularly in the band gap ($E_g$) region [13]. When $\alpha$ spectra calculated within generalized gradient approximation (GGA) [18] using high-density $k$ mesh are blue-shifted toward higher energy to compensate the underestimated $E_g$ contribution in GGA, the DFT and experimental spectra show remarkable agreement [13]. Later, it was suggested to shift the GGA-calculated $\alpha$ spectra using $E_g$ values estimated from hybrid-functional DFT calculations [19]. However, the validity of such a method has not been discussed properly.

In this study, to realize a highly accurate prediction method of material $\alpha$ spectra, we have developed a quite general DFT calculation scheme, in which $\alpha$ spectra calculated by GGA using very high $k$-mesh density are blue-shifted toward more accurate energy scale determined by a hybrid functional (HSE06) [20,21] while performing the amplitude correction simultaneously based on sum rule [22]. A key feature of our method is the combination of GGA within the Perdew-Burke-Ernzerhof scheme (PBE) [18] with HSE06 and sum rule and, from this PHS approach (PBE+HSE06+Sum rule), the dielectric function ($\varepsilon = \varepsilon_1 - i\varepsilon_2$) and optical constants (refractive index $n$, extinction coefficient $k$ and $\alpha$) are readily obtained in a consistent manner. As a result, we find that the $\alpha$ spectra of representative solar cells materials [GaAs, CdTe, InP, CuInSe$_2$ (CISe), and Cu$_2$ZnGeSe$_4$ (CZGSe)], calculated by the developed PHS method, show remarkable agreement with the experimental spectra in a wide $\alpha$ range of $10^2 \sim 10^6$ cm$^{-1}$. The agreement of the $\alpha$ spectra observed in this study is far better than those obtained in general methods based on GGA, hybrid functional and GW calculations. Our approach provides an ideal method for accurate prediction of overall material optical properties, which can be incorporated directly into optical device simulations.

## 2. PHS method

Figure 1 explains the calculation procedure of the PHS method developed in this study. Here, as an example, the calculation of a GaAs dielectric function is shown. In our approach, the $\varepsilon_2$ spectrum of the dielectric function is calculated first using very high $k$-mesh density within GGA-PBE. As known well [23], $E_g$ is seriously underestimated when DFT calculations are performed within PBE. To compensate this underestimated $E_g$ contribution, the calculated PBE $\varepsilon_2$ spectrum ($\varepsilon_{2,\text{PBE}}$) is blue-shifted. In this energy shift, the $E_g$ correction value ($\Delta E_g$) is determined according to $\Delta E_g = E_{g,\text{HSE}} - E_{g,\text{PBE}}$, where $E_{g,\text{HSE}}$ and $E_{g,\text{PBE}}$ represent the $E_g$ values estimated from HSE06 and PBE, respectively.

When the $\varepsilon_2$ spectrum is shifted toward higher energy, however, it is necessary to satisfy sum rule [22], given by

$$\int E\varepsilon_2(E)dE = \text{const.} \tag{1}$$

If the $\varepsilon_2$ spectrum is shifted toward higher energy by $\Delta E_g$, sum rule requires that

$$\int E\varepsilon_2(E)dE = \int (E+\Delta E_g)\varepsilon_{2,\text{Shift}}(E+\Delta E_g)dE, \tag{2}$$

where $\varepsilon_{2,\text{Shift}}$ shows the shifted $\varepsilon_2$ spectrum. In order for Eq. (2) to be satisfied, the amplitude of $\varepsilon_{2,\text{Shift}}$ needs to be reduced by a factor of $f = E/(E + \Delta E_g)$ [16] and, by applying this principle, we obtain $\varepsilon_{2,\text{Shift}}(E + \Delta E_g) = f\varepsilon_2(E)$. If we convert the $E$ axis of this equation using $E + \Delta E_g \rightarrow E$ (i.e., $E \rightarrow E - \Delta E_g$) and assume that $\varepsilon_2(E) = \varepsilon_{2,\text{PBE}}(E)$, the $\varepsilon_2$ spectrum of the PHS method [$\varepsilon_{2,\text{PHS}}(E)$] is calculated by setting $\varepsilon_{2,\text{PHS}}(E) = \varepsilon_{2,\text{Shift}}(E)$ as follows:

$$\varepsilon_{2,\text{PHS}}(E) = \frac{E - \Delta E_g}{E}\varepsilon_{2,\text{PBE}}(E - \Delta E_g). \tag{3}$$

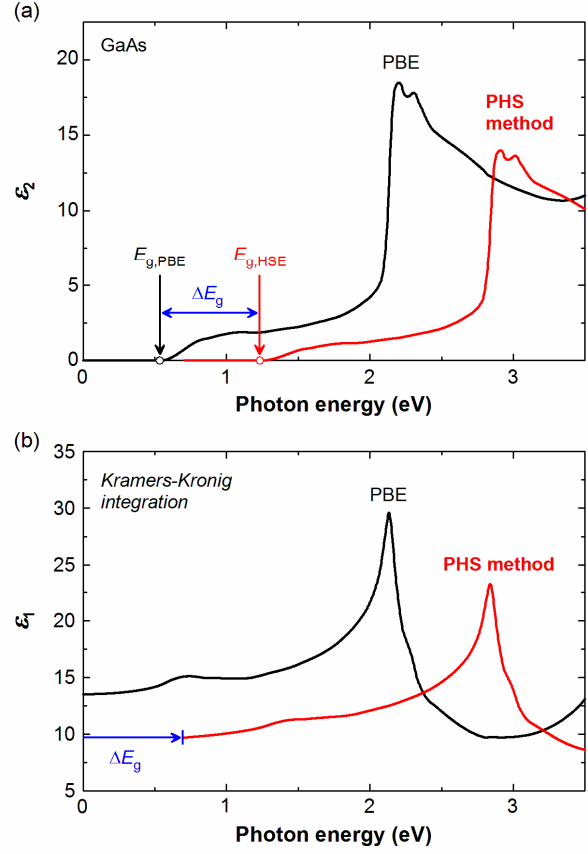

Fig. 1 Calculation of (a) $\varepsilon_2$ spectrum and (b) $\varepsilon_1$ spectrum for GaAs based on the PHS method. In (a), the determination of a $E_g$ correction value ($\Delta E_g = E_{g,HSE} - E_{g,PBE}$) is indicated. The $\varepsilon_1$ spectrum is obtained from Kramers-Kronig integration of (a).

In Fig. 1(a), the calculation results of $\varepsilon_{2,PHS}$ and $\varepsilon_{2,PBE}$ are shown. The $\varepsilon_1$ contribution of the PHS method ($\varepsilon_{1,PHS}$) can then be obtained from Kramers-Kronig integration [22]:

$$\varepsilon_{1,PHS}(E) = 1 + \frac{2}{\pi} P \int_0^\infty \frac{E' \varepsilon_{2,PHS}(E')}{E'^2 - E^2} dE'. \quad (4)$$

The results of $\varepsilon_{1,PHS}$ and $\varepsilon_{1,PBE}$ are compared in Fig. 1(b). From ($\varepsilon_{1,PHS}$, $\varepsilon_{2,PHS}$), $n_{PHS}$ and $k_{PHS}$ are determined by conventional formula:

$$n_{PHS} = \{[\varepsilon_{1,PHS} + (\varepsilon_{1,PHS}^2 + \varepsilon_{2,PHS}^2)^{1/2}]/2\}^{1/2}, \quad (5)$$

$$k_{PHS} = \{[-\varepsilon_{1,PHS} + (\varepsilon_{1,PHS}^2 + \varepsilon_{2,PHS}^2)^{1/2}]/2\}^{1/2}. \quad (6)$$

Finally, the $\alpha$ spectrum of the PHS method is deduced as $\alpha_{PHS} = 4\pi k_{PHS}/\lambda$.

## 3. DFT calculation

The DFT calculations were performed using Advance/PHASE and the Vienna *Ab initio* Simulation Package (VASP) [24]. For the calculations of GGA within PBE, the Advance/PHASE software was employed, while the VASP software was applied for HSE06 calculations. For the DFT calculations of zincblende crystals (GaAs, CdTe, InP), two-atom primitive cells were used, while eight-atom primitive cells were employed for CISe and CZGSe. The structural optimization of all the crystals was made by HSE06 using a plane-wave cutoff energy of 455 eV and the structures obtained from this procedure were applied for all the optical function calculations implemented by PBE and HSE06.

The optical-function calculations using PBE were made based on a method developed by Kageshima et al [25]. In this calculation, plane-wave ultrasoft pseudopotential and tetrahedron methods were adopted. In addition, for Cu-containing compounds (CISe and CZGSe), the onsite Coulomb interaction was considered for the Cu $3d$ state [26] with an effective energy of $U_{eff}$ = 3 eV. Unless otherwise noted, for the PBE calculations, we used a highly dense $30 \times 30 \times 30$ $k$ mesh for GaAs, CdTe and InP, whereas a $16 \times 16 \times 16$ $k$ mesh was employed for CISe and CZGSe [13,14]. The above $k$ mesh densities were chosen so that the $k$ mesh density in the $k$ space becomes less than 0.1 Å$^{-1}$. We previously confirmed that this $k$ mesh density provides satisfactory agreement with experimental results [13,14].

The optical-function calculation based on HSE06 was implemented for GaAs using a less dense $16 \times 16 \times 16$ $k$ mesh, compared with PBE, due to the extensive calculation cost of HSE06. In this case, the $k$-space integration was made based on a Γ-centered Monkhorst-Pack method and the obtained optical spectrum was broadened with a complex shift of $\eta$ = 0.1.

## 4. Results and Discussion

Figure 2 shows (a) the $\varepsilon_2$ spectra and (b) the $\varepsilon_1$ spectra of GaAs obtained from experiment (open circles) and the DFT calculations (solid lines). For the DFT results, those obtained applying PBE, HSE06 and the PHS method are shown. The experimental spectrum was taken from Ref. [17]. Since $E_g$ is seriously underestimated in PBE, the whole PBE spectrum is red-shifted, compared with the experimental spectrum. In contrast, the HSE06 calculation provides a better fitting to the experimental spectrum and the optical transition energies observed in the experimental spectrum are reproduced well, as reported previously [9]. When the PHS method is applied, we obtain the $\varepsilon_2$ spectrum shown by the red line in Fig. 2(a) ($\Delta E_g = 0.705$ eV), which provides the better overall agreement with the experimental spectrum, compared with the HSE06 result. As shown in Fig. 2(b), the $\varepsilon_1$ spectrum calculated by the PHS approach also provides satisfactory agreement with the experimental result.

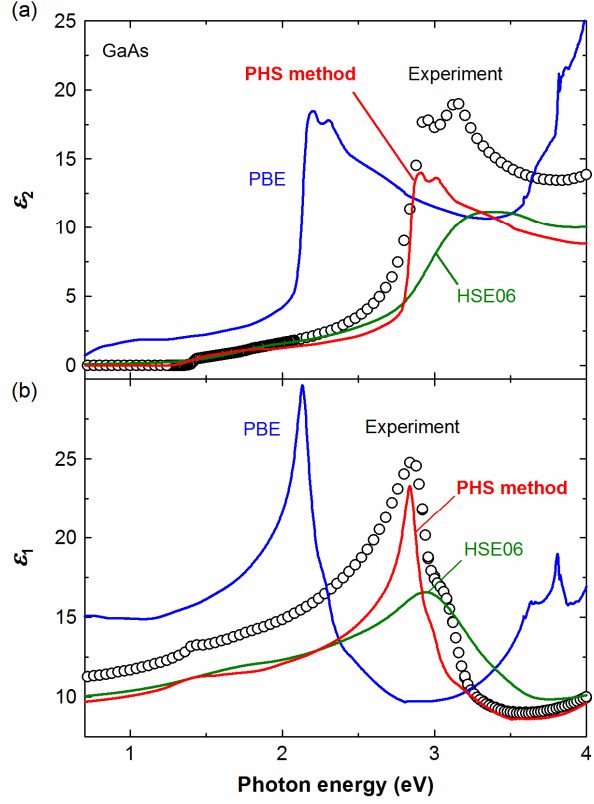

Fig. 2. (a) $\varepsilon_2$ spectra and (b) $\varepsilon_1$ spectra of GaAs obtained from experiment (open circles) and theoretical DFT calculations (solid lines). For the calculations, the results determined by PBE, HSE06, and the developed PHS method are shown. The experimental data were taken from Ref. [17].

Figure 3 shows (a) the comparison of the GaAs $\alpha$ spectra obtained from experiment [17] and the PHS method and (b) the GaAs $\alpha$ spectra calculated by the PHS, HSE06 and GW methods. The $\alpha$ spectrum estimated from GW is adopted from Ref. [8]. In Fig. 3(a), the $\alpha$ spectra calculated by the PHS method using different *k*-mesh densities are shown. It can be seen that the DFT calculation using a very high mesh density is vital for the accurate calculation of the $\alpha$ spectrum in logarithmic scale. This is based on the fact that the light absorption in the $E_g$ region is highly localized near the Γ point in the Brillouin zones in conventional tetragonal-based semiconductors [13,27] and the precise *k*-space calculation particularly around the Γ point is necessary to reproduce the band-edge optical transition accurately [13]. As a result, the agreement with the experimental $\alpha$ improves significantly as the *k*-mesh density is increased. As mentioned above, sum rule is incorporated in our method. When the $\varepsilon_2$ spectrum calculated by PBE is shifted toward higher energy without considering sum rule, $\alpha$ is overestimated notably in the $E_g$ region (supplementary material, Fig. 1). Accordingly, the incorporation of sum rule and the following Kramers-Kronig integration are essential.

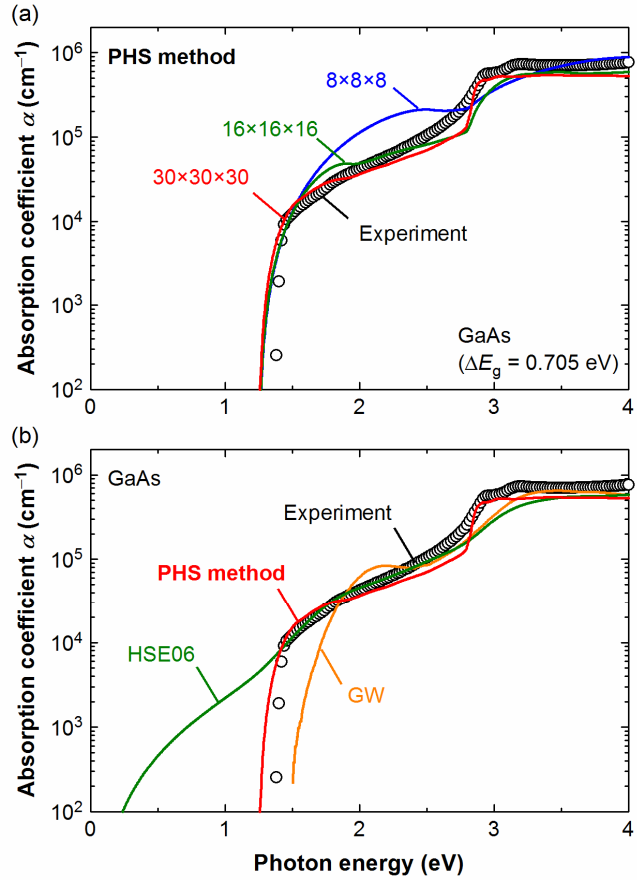

Fig. 3. (a) GaAs $\alpha$ spectra calculated by the PHS method using different *k*-mesh densities and (b) comparison of GaAs $\alpha$ spectra calculated by the PHS, HSE06 and GW methods. The experimental result (open circles) is taken from Ref. [17] and the GW spectrum in (b) is adopted from Ref. [8]. The $\Delta E_g$ represents the energy-shift value of the PBE spectra ($\Delta E_g = E_{g,\text{HSE}} - E_{g,\text{PBE}}$).

As confirmed from Fig. 3(b), the agreement with the experimental result improves drastically when the developed PHS method is applied. Although the $\varepsilon_2$ spectrum calculated by HSE06 shows a good agreement in linear scale [see Fig. 2(a)], the agreement is poor in logarithmic $\alpha$ scale. The notable light absorption observed even below $E_g$ in HSE06 is caused by smoothening of the calculated $\varepsilon_2$ spectrum and such an artifact can be eliminated when the broadening parameter ($\eta$) is set to zero (supplementary material, Fig. 2). On the other hand, the GW spectrum reproduces the overall experimental spectrum well. Nevertheless, the agreement in the $E_g$ region is quite inferior in the GW spectrum, if compared with the PHS method. Moreover, the calculation cost of the GW method is quite high and the application of high-density $k$ mesh calculation is generally limited in the case of GW.

As a result, only the developed PHS method provides a satisfactory agreement with the experimental $\alpha$ spectrum in logarithmic scale. It should be emphasized that the band-edge $\alpha$ has a significant impact on the operation of solar cell devices [16,28] as the photocarrier collection becomes more difficult in the near-$E_g$ region due to the increase in the light penetration depth ($d_p = 1/\alpha$). Thus, accurate prediction of the band-edge $\alpha$ is of paramount importance particularly when DFT spectra are applied directly to optical device simulations.

Furthermore, our PHS approach realizes the fast calculation of $\alpha$ spectra, as the high-density $k$ mesh calculation is implemented using a rather simple PBE approximation. In our calculations, for example, the calculation time of the GaAs dielectric function using PBE is 0.1 s per $k$-point (45 min in total for a 30 × 30 × 30 $k$ mesh), while a similar calculation using HSE06 results in 4.8 s per $k$-point (325 min in total for a less dense 16 × 16 × 16 $k$ mesh). Accordingly, our PHS method has clear advantages over more general HSE06 and GW methods in terms of accuracy and calculation cost. It should be noted that, when the $\alpha$ spectrum calculated by PBE is simply blue-shifted by $\Delta E_g$, we obtain a spectrum similar to the one calculated from the PHS method (supplementary material, Fig. 3). Thus, although the rigorous approach of the PHS method is preferable, the shifted PBE $\alpha$ spectrum could also be adopted [13].

We have applied the PHS method for the $\alpha$ calculation of other solar cell materials. Figure 4 shows the $\alpha$ spectra of (a) CdTe, (b) InP, (c) CISe and (d) CZGSe, obtained from experiment [14,17,29] and the calculations using the PHS method. The $\Delta E_g$ values determined from the calculations are also indicated. In Fig. 4, all the calculated results show remarkable overall agreement with the experimental spectra, confirming the universality of our PHS approach.

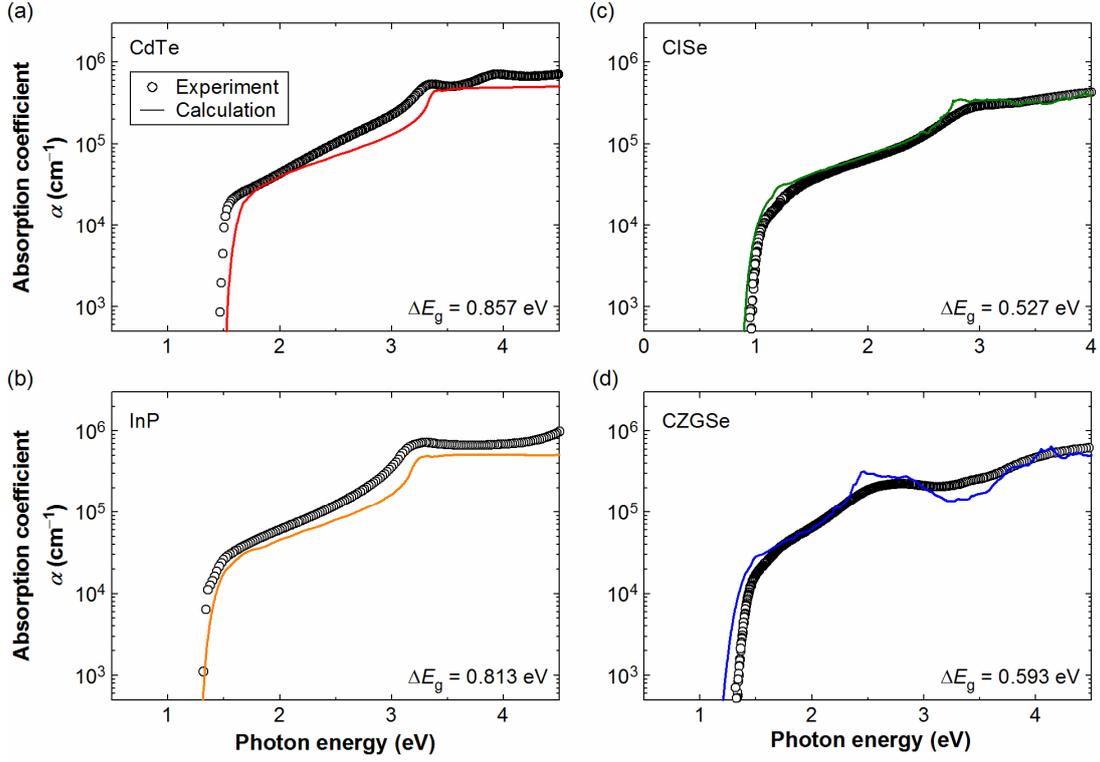

Fig. 4. $\alpha$ spectra of (a) CdTe, (b) InP, (c) CISe and (d) CZGSe, obtained from experiment (open circles) and the calculations using the PHS method (solid lines). The $\Delta E_g$ values of each spectrum are also indicated. The experimental data were adopted from Refs. [14], [17] and [29].

To justify our method further, we have calculated the band structures of the solar cell materials. Figure 5 compares the band structures of (a) CdTe, (b) InP, (c) CISe and (d) CZGSe, obtained from PBE and HSE06. In the PBE results of Fig. 5, all the conduction band positions were shifted upward by $\Delta E_g$ so that $E_g$ becomes consistent with that obtained from HSE06 (scissor operation [30]). As known well [31], the underestimation of $E_g$ within PBE originates from the assumption that only non-interacting single particle is considered in the $E_g$ estimation; however, the variations of individual conduction and valence bands deduced from PBE are still accurate. When the PBE conduction bands are shifted, therefore, all the PBE bands show excellent agreement with the bands approximated by HSE06 [13]. A similar good agreement has also been observed for GaAs (supplementary material, Fig. 4). Since the optical transitions are derived essentially from the band structures, the result of Fig. 5 verifies that the underestimated $E_g$ contribution observed in the PBE spectra can be corrected by simply shifting the PBE spectra toward higher energies using $\Delta E_g$. In other words, the validity of our PHS method can be confirmed by simply comparing the conduction-band-shifted PBE band structure with the HSE06 band structure. Since the band-edge light absorption of tetragonal-based semiconductors is determined primarily by the optical transition from the first valence band to the first conduction band [27], the agreement near the valence band maximum and conduction band minimum is particularly important.

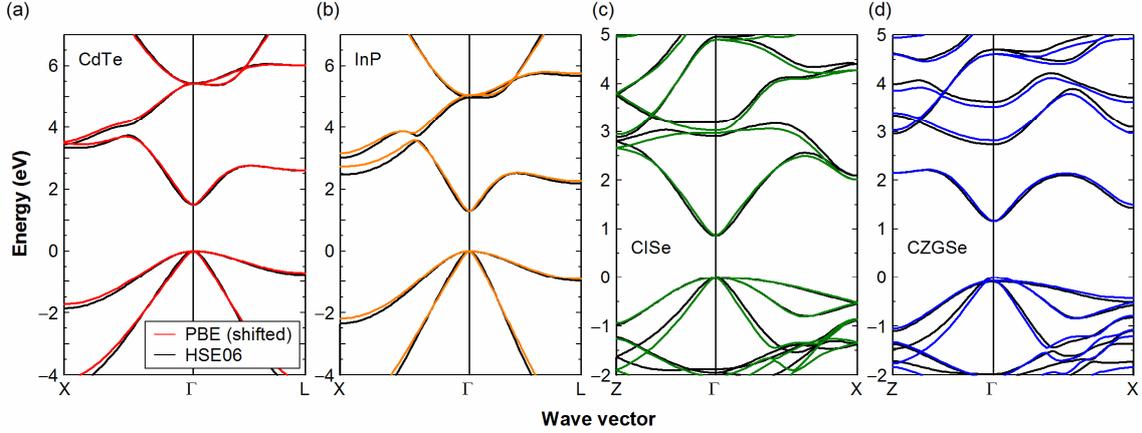

Fig. 5. Band structures of (a) CdTe, (b) InP, (c) CISe and (d) CZGSe calculated using PBE and HSE06. In the PBE results, all the conduction band positions were shifted upward by $\Delta E_g$.

## 5. Conclusion

We have developed a new DFT approach that can accurately predict material $\alpha$ spectra in logarithmic scale. In this method, the $\varepsilon_2$ spectrum calculated from the PBE functional using very high-$k$ mesh density is blue-shifted and the underestimated $E_g$ contribution in PBE is corrected using the energy scale determined by HSE06 calculation, while the $\varepsilon_2$ amplitude is corrected by applying sum rule. We have applied the developed method for the $\alpha$ calculations of five solar cell materials (GaAs, InP, CdTe, CuInSe$_2$ and Cu$_2$ZnGeSe$_4$) and the $\alpha$ spectra calculated by our method provide remarkable agreement with those observed experimentally. Our scheme, which is superior to the HSE06 and GW calculations, has clear advantages in accuracy and calculation cost and allows the direct application of calculated DFT optical spectra to various optical device simulations.

**Acknowledgement**

This work was supported by JSPS KAKENHI Grant Number JP19H02167.

**Data availability**

The raw data required to reproduce these findings are available. The processed data required to reproduce these findings are also available.

**Reference List**

[1] Y. Hinuma, T. Hatakeyama, Y. Kumagai, L.A. Burton, H. Sato, Y. Muraba, S. Iimura, H. Hiramatsu, I. Tanaka, H. Hosono, F. Oba, Nat. Commun. 7 (2016) 11962.

[2] V.L. Shaposhnikov, A.V. Krivosheeva, V.E. Borisenko, J.-L. Lazzari, F. Arnaud d'Avitaya, Phys. Rev. B 85 (2012) 205201.

[3] W. Meng, B. Saparov, F. Hong, J. Wang, D.B. Mitzi, Y. Yan, Chem. Mater. 28 (2016) 821-829.

[4] M. Kato, T. Fujiseki, T. Miyadera, T. Sugita, S. Fujimoto, M. Tamakoshi, M. Chikamatsu, H. Fujiwara, J. Appl. Phys. 121 (2017) 115501.

[5] M. Gajdoš, K. Hummer, G. Kresse, J. Furthmüller, F. Bechstedt, Phys. Rev. B 73 (2006) 045112.

[6] C. Ambrosch-Draxl, J.O. Sofo, Comput. Phys. 175 (2006) 1-14.

[7] F. Kootstra, P.L. de Boeij, J.G. Snijders, Phys. Rev. B 62 (2000) 7071-7083.

[8] S. Botti, F. Sottile, N. Vast, V. Olevano, L. Reining, H.-C. Weissker, A. Rubio, G. Onida, R.D. Sole, R.W. Godby, Phys. Rev. B 69 (2004) 155112.

[9] J. Paier, M. Marsman, G. Kresse, Phys. Rev. B 78 (2008) 121201.

[10] H. Fujiwara, M. Kato, M. Tamakoshi, T. Miyadera, M. Chikamatsu, Phys. Status Solidi A 215 (2018) 1700730.

[11] J. Paier, R. Asahi, A. Nagoya G. Kresse, Phys. Rev. B 79 (2009) 115126.

[12] C. Persson, J. Appl. Phys. 107 (2010) 053710.

[13] M. Nishiwaki, K. Nagaya, M. Kato, S. Fujimoto, H. Tampo, T. Miyadera, M. Chikamatsu, H. Shibata, H. Fujiwara, Phys. Rev. Materials 2 (2018) 085404.

[14] K. Nagaya, S. Fujimoto, H. Tampo, S. Kim, M. Nishiwaki, Y. Nishigaki, M. Kato, H. Shibata, H. Fujiwara, Appl. Phys. Lett. 113 (2018) 093901.

[15] E.M. Chen, L. Williams, A. Olvera, C. Zhang, M. Zhang, G. Shi, J.T. Heron, L. Qi, L.J. Guo, E. Kioupakis, P.F.P. Poudeu, Chem. Sci. 9 (2018) 5405-5414.

[16] A. Nakane, H. Tampo, M. Tamakoshi, S. Fujimoto, K.M. Kim, S. Kim, H. Shibata, S. Niki, H. Fujiwara, J. Appl. Phys. 120 (2016) 064505.

[17] H. Fujiwara, R.W. Collins, editors, *Spectroscopic Ellipsometry for Photovoltaics: Volume 2: Applications and Optical Data of Solar Cell Materials*, Springer, Cham, 2018.

[18] J.P. Perdew, K. Burke, M. Ernzerhof, Phys. Rev. Lett. 77 (1996) 3865-3868.

[19] J. Wang, H. Chen, S.-H. Wei, W.-J. Yin, Adv. Mater. 31 (2019) 1806593.

[20] J. Heyd, G.E. Scuseria, M. Ernzerhof, J. Chem. Phys. 118 (2003) 8207-8215.

[21] J. Heyd G.E. Scuseria, M. Ernzerhof, J. Chem. Phys. 124 (2006) 219906.

[22] M.P. Marder, Condensed Matter Physics, Wiley, Hoboken, 2010.

[23] M. Marsman, J. Paier, A. Stroppa, G. Kresse, J. Phys.: Condens. Matter 20 (2008), 064201.

[24] G. Kresse, J. Furthmüller, Phys. Rev. B 54 (1996) 11169-11186.

[25] H. Kageshima, K. Shiraishi, Phys. Rev. B 56 (1997) 14985-14992.

[26] Y. Zhang, X. Yuan, X. Sun, B.-C. Shih, P. Zhang, W. Zhang, Phys. Rev. B 84 (2011) 075127.

[27] M. Kato, M. Nishiwaki, H. Fujiwara, 2019, https://arxiv.org/abs/1906.03005.

[28] M. Shirayama, H. Kadowaki, T. Miyadera, T. Sugita, M. Tamakoshi, M. Kato, T. Fujiseki, D. Murata, S. Hara, T.N. Murakami, S. Fujimoto, M. Chikamatsu, H. Fujiwara, Phys. Rev. Applied 5 (2016) 014012.

[29] S. Minoura, K. Kodera, T. Maekawa, K. Miyazaki, S. Niki, H. Fujiwara, J. Appl. Phys. 113 (2013) 063505.

[30] F. Gygi, A. Baldereschi, Phys. Rev. Lett. 62 (1989) 2160-2163.

[31] F. Bechstedt, *Many-Body Approach to Electronic Excitations: Concepts and Applications*, Springer, Heidelberg, 2015.

**Highly accurate prediction of material optical properties based on density functional theory**

Mitsutoshi Nishiwaki and Hiroyuki Fujiwara

*Department of Electrical, Electronic and Computer Engineering, Gifu University, 1-1 Yanagido, Gifu 501-1193, Japan*

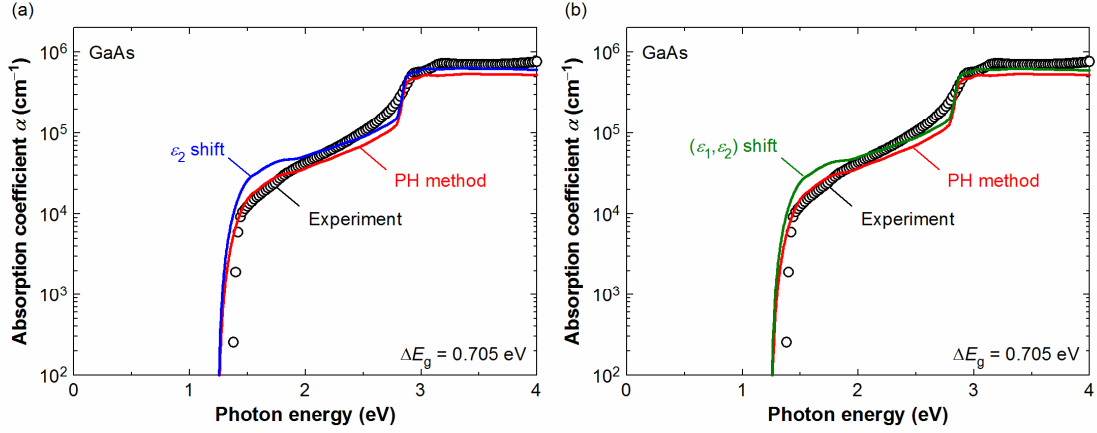

Fig. S1. DFT calculation results obtained from the simple $\varepsilon_2$ spectral shift without incorporating sum rule: (a) with Kramers-Kronig integration and (b) without Kramers-Kronig integration. The solid lines show calculated $\alpha$ spectra of GaAs, whereas the open circles show the experimental result. The red lines indicate the DFT spectrum estimated from the PHS method. In the DFT calculation of (a), the PBE $\varepsilon_2$ spectrum is simply shifted toward higher energy by $\Delta E_g$ without the amplitude correction (i.e., without sum rule) and the corresponding $\varepsilon_1$ spectrum is obtained from Kramers-Kronig integration. Finally, from the ($\varepsilon_1$, $\varepsilon_2$) pair, the $\alpha$ spectrum [blue line in (a)] is obtained. In the case of (b), both $\varepsilon_1$ and $\varepsilon_2$ spectra obtained from the PBE calculation are shifted toward higher energy and the $\alpha$ spectrum is calculated directly from these shifted $\varepsilon_1$ and $\varepsilon_2$ spectra [green line in (b)].

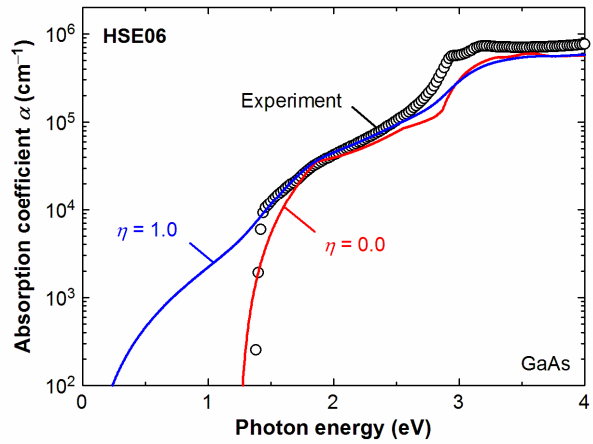

Fig. S2. GaAs $\alpha$ spectra calculated by HSE06 using different broadening parameters ($\eta$). When $\eta$ was set to zero, the corresponding $\varepsilon_1$ spectrum was calculated from the $\varepsilon_2$ spectrum by applying Kramers-Kronig integration.

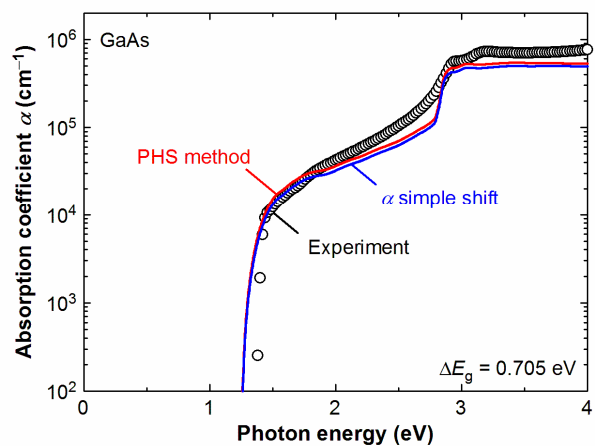

Fig. S3. Comparison of a DFT $\alpha$ spectrum obtained from the simple shifting of the PBE-$\alpha$ spectrum toward higher energy with the one obtained from the PHS method. Both results are almost identical.

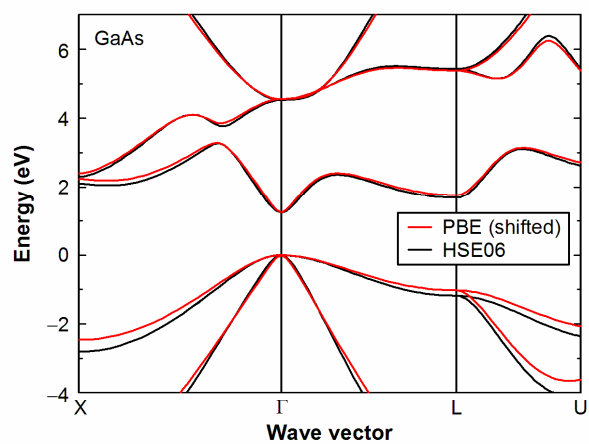

Fig. S4. Band structures of GaAs calculated using PBE and HSE06. In the PBE results, all the conduction band positions were shifted upward by $\Delta E_g$.

16